\documentclass[prl,twocolumn,showpacs,preprintnumbers,amsmath,amssymb]{revtex4}

\usepackage{graphicx}
\usepackage{dcolumn}
\usepackage{bm}
\usepackage{subfigure}

\newcommand{\degree}{\ensuremath{^\circ}}

\begin{document}


\title{Origin of magnetism and quasiparticles properties in Cr-doped TiO$_2$}

\author{F. Da Pieve$^{1,2,3}$, S. Di Matteo$^{4}$, T. Rangel$^{2,3}$, M. Giantomassi$^{2,3}$, D. Lamoen$^{1}$, G.-M. Rignanese$^{2,3}$ and X. Gonze$^{2,3}$}
\affiliation{
$^1$ EMAT, University of Antwerp, Groenenborgerlaan 171, B-2020 Antwerp, Belgium\\
$^2$ Institute of Condensed Matter and Nanoscience, Universit\'e Catholique de Louvain, Chemin des \'Etoiles 8, bte L7.03.01, (B) 1348 Louvain-la-Neuve, Belgium\\
$^3$ ETSF, European Theoretical Spectroscopy Facility\\
$^4$ Groupe Th\'eorie D\'epartement Mat\'eriaux et Nanoscience Institut de Physique de Rennes, UMR UR1-CNRS 6251, Universit\'e de Rennes 1, F-35042 Rennes Cedex, France
}

\date{\today}

\begin{abstract}{Combining LSDA+$U$ and an analysis of superexchange interactions beyond DFT, we describe the magnetic ground states in rutile and anatase Cr-doped TiO$_2$. In parallel, we correct our LSDA+$U$ ground state through GW corrections ($GW$@LSDA+$U$) that reproduce the position of impurity states and the band gaps in satisfying agreement with experiments. Because of the different topological coordinations of Cr-Cr bonds in the ground states of rutile and anatase, superexchange interactions induce either ferromagnetic or antiferromagnetic couplings of Cr ions. In Cr-doped anatase, this interaction leads to a new mechanism which stabilizes a ferromagnetic ground state, in keeping with experimental evidence, without the need to invoke F-center exchange.}

\end{abstract}

\maketitle



Understanding the physics of pure and doped transition-metal oxides (TMO), in order to have reliable predictions of their ground and excited states, represents a major challenge in fundamental and applied research today. The main difficulty for an {\it ab-initio} description of the TMO wavefunctions stems from the partly localized character of their $3d$-electron states, which often leads to deep interplays of charge, spin and orbital degrees of freedom \cite{science288}. For this reason, Density Functional Theory (DFT), within its traditional functionals, fails to describe their electronic and magnetic properties \cite{secondo}, as Kohn-Sham wave functions are too delocalized. In order to cure such a drawback, alternative schemes like hybrid functionals \cite{janesko}, SIC \cite{droghetti} and LDA+$U$ \cite{method} have been used in recent years. The latter, in particular, due to the strong on-site Coulomb repulsion, leads to a more pronounced localization of $3d$ wave-functions that better represents their behaviour and can therefore be used as a zeroth-order for subsequent many-body corrections ($GW$@LSDA+$U$ approach \cite{method,other1,other2}).

The recent discovery of room temperature ferromagnetism (FM) in TM-doped TiO$_2$ \cite{science291} has triggered an enormous experimental and theoretical effort to understand the magnetic properties of a class of compounds (TiO$_2$, ZnO, SnO) that show correlation effects when doped with TM impurities. Though F-center exchange is often invoked to justify FM in these and analogous systems \cite{coey}, recent {\it ab-initio} calculations \cite{freeman,jannotti} point to its exclusion, so that the actual physical mechanism remains an open question \cite{appl,appld,thin,droub,prl95,prb73}. Theoretical results with advanced functionals on TM-doped TiO$_2$ \cite{dival,kyangmio} allow to qualitatively reproduce band gaps, but do not explain FM.

In this work we describe magnetic ground states quantitatively by evaluating superexchange (SE) corrections to the LSDA+$U$ zeroth-order magnetic energies and discuss their relevance for the stability of magnetic ground states in connection with the topology of the host lattice. Our approach goes beyond DFT (that does not consider dynamic fluctuations) and current implementations of Dynamical Mean-Field Theory, limited to atomic dynamical correlations \cite{dmft}. The correctness of the zeroth-order LSDA+$U$ calculations is confirmed by our $GW$@LSDA+$U$ quasiparticle results, reproducing photoemission experiments with satisfying agreement. 
With this calculation scheme, we perform a systematic analysis of rutile and anatase Cr-doped TiO$_2$ and confirm that F-center exchange is not active in Cr-doped TiO$_2$, as O vacancies are F$^{++}$-centers. Moreover, the investigation of SE interactions highlights a strong interplay between the topology of the local structure, the presence of an oxygen vacancy and the orbital occupation of impurity levels, leading to the conclusion that such oxygen vacancies can boost FM via a different physical mechanism than F-center exchange. The influence of oxygen vacancies on the positions of impurity states is also investigated. {\it Mutatis mutandis}, our approach can be adapted to a large class of doped TMO and it paves the way for a critical re-examination of magnetism in all those compounds where $t_{2g}$ levels play a major role and a simplistic application of the Goodenough rules \cite{goodenough} is not appropriate.


{\it Calculation framework.}
Calculations are performed within the PAW scheme using the ABINIT package \cite{abi}. Ti and Cr semicore states are included in the valence. We used supercells with 48 atoms sampling the Brillouin zone with 2$\times$2$\times$3 and 2$\times$2$\times$2 Monkhorst-Pack k-mesh for rutile and anatase. Structural degrees of freedom are optimized by minimizing quantum mechanical forces and total energy. $G_0W_0$ calculations are performed using the Godby-Needs plasmon-pole approximation \cite{godby}, and with 2000 bands for the Green's function. SE  calculations are described below.
 
\begin{figure}[b]
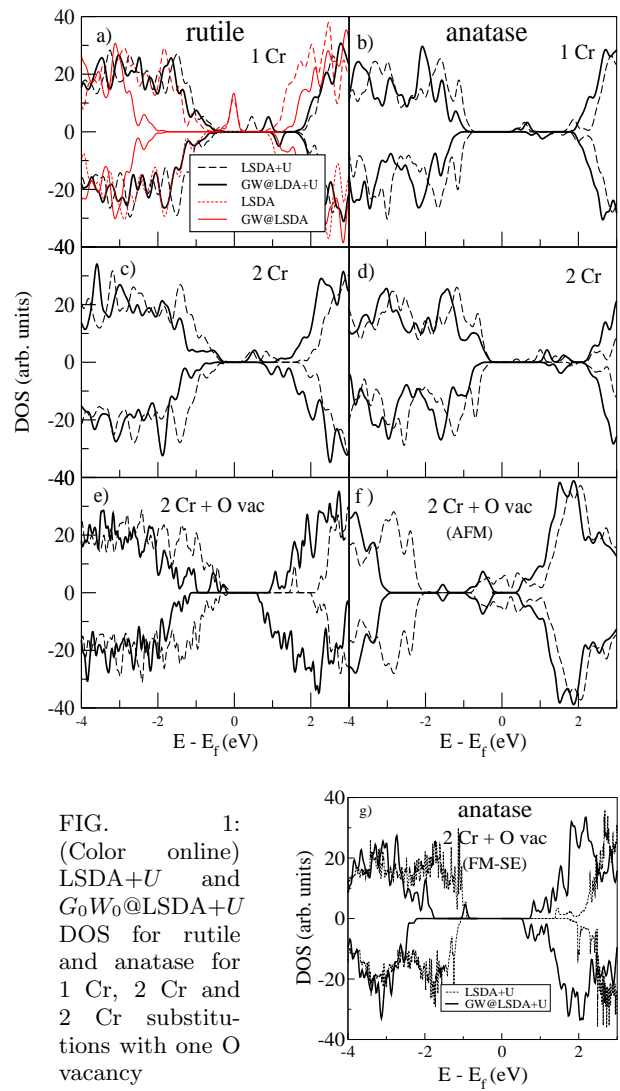

\begin{center}
\includegraphics[clip=,height=10.0cm,width=0.45\textwidth]{FigDOS.eps}
\end{center}
\parbox[b]{.28\columnwidth}{

\caption{(Color online) LSDA+$U$ and $G_0W_0$@LSDA+$U$ DOS for rutile and anatase for 1 Cr, 2 Cr and 2 Cr substitutions with one O vacancy}
\label{conf2}
}
\parbox[b]{0.50\columnwidth}{
\hspace*{1.9em}\includegraphics[clip=,height=4.0cm,width=0.24\textwidth]{Figanatonly.eps}}\hfill
\end{figure}

At first, we confirm that LSDA is a poor zeroth-order approach for TMO: a direct application of $G_0W_0$ on the LSDA half-metallic state does not open a gap and the only effect of the $G_0W_0$ self energy is to push down valence states with respect to the conduction band [see Fig. \ref{conf2}(a)]. Since the states near the Fermi level are largely of $t_{2g}$ character, we describe their partly localized behaviour by LSDA+$U$. We consider $U$=5.0 (5.2) eV for Ti (Cr) $3d$ states, from the literature \cite{bouquet}, and use the values of Hund's coupling $J$ as obtained from LSDA: 0.80 (0.84) eV for Ti (Cr), as $J$ is almost unsensitive to screening effects. The LSDA+$U$ and $G_0W_0$@LSDA+$U$ density of states (DOS) for the cases of one Cr substitution, two Cr substitutions, and two Cr in combination with an O vacancy are shown in Fig. \ref{conf2} for the most stable configurations both in rutile and anatase. We also show the FM case for anatase with 2 Cr and 1 O vacancy. Our calculations correspond to Cr concentrations of 6.25\% and 12.5\%. The band gaps and positions of impurity states are summarized in Table I.

\begin{table}
\caption{\label{tab:table1} Band gaps of majority and minority spins for rutile (r) and anatase (a) in LSDA+$U$ and $G_0W_0$@LSDA+$U$. Positions of impurity states of given spin polarization with respect to the top of the valence band are shown in parenthesis. In bold, values to be compared with experimental data (see text). All values are in eV.}
\begin{ruledtabular}
\begin{tabular}{lcc}
 & LSDA+$U$ ($\uparrow$; $\downarrow$) & $G_0W_0$ ($\uparrow$; $\downarrow$) \\
\hline
1Cr (r) & 2.0 (0.6); 2.0 & 1.9 (0.9); 1.9 (1.1)    \\
2Cr (r)   & 1.8 (0.5); 1.7  & 1.3 (0.6); 1.0  \\
2Cr-O$_v$ (r) & 2.2 (1.5); 2.2  & {\bf 2.0 (0.2)}; {\bf 1.9 }   \\
1Cr (a)  & 2.3 (1.0); 2.4 & 2.7 (1.1); 2.7 (1.4) \\
2Cr (a)  & 0.5; 1.8  & 2.4 (1.2); 2.5 (1.3)   \\
2Cr-O$_v$ (a)  & 1.6; 1.6  & {\bf 2.5 (0.7)}; {\bf 2.7} \\
\end{tabular}
\end{ruledtabular}
\end{table}

{\it LSDA+$U$ results.} The LSDA+$U$ functional opens a gap at the Fermi energy for both rutile and anatase phases (with and without O vacancy). The band gaps and positions of impurity states are summarized in Table I and will be compared with available experiments later in the text. Here we focus ourselves on discussing the interplay between local structural, electronic and magnetic properties. The DOS of the impurity states is mainly composed of Cr $t_{2g}$ states. Cr-doping favors the formation of an O vacancy, as the formation energy is lowered with respect to the undoped cases by 1.2 and 1.6 eV for rutile and anatase, respectively. For the lattices with 2 Cr ions, such an O vacancy is found to be more stable at a bridging position between the two impurities, in both lattices. 
Interestingly, the most stable configurations in rutile and anatase, both with and without oxygen vacancies, are characterized by a topological difference in the relative position of the two Cr ions: nearest neighbors (NN) in rutile, connected along the tetragonal axis (bond-angle Cr-O-Cr $\sim 90\degree$) and next nearest neighbors (NNN) in anatase, connected along a direction perpendicular to the tetragonal axis (bond angle $\sim 180\degree$). This result is also valid for LSDA calculations and is probably favoured by the relative closeness of NNN in the anatase lattice (3.8 \AA) compared to rutile (4.6 \AA). 
Furthermore, we found that the electron charge density at the O vacancy is negligible (see Supplementary Material), i.e., the vacancy is a F$^{++}$-center \cite{serpone}, thereby suggesting that F-center exchange \cite{coey} is not active in this case and that the origin of magnetism should be looked for elsewhere.

Spin polarization mainly occurs at the Cr site ($\sim 99\%$). Without oxygen vacancies, we obtain a total magnetic moment of 2 $\mu_B$/supercell, i.e., the two $t_{2g}$ electrons in a triplet configuration. Removal of one neutral O atom in the supercell leaves two extra electrons that reduce two Cr$^{4+}$ to Cr$^{3+}$ ions leading to a bigger average magnetic moment of 3 $\mu_B$/Cr atom, in keeping with the values 2.6-2.9 $\mu_B$/Cr atom observed in Refs. \cite{appl,nhong}. 
We also performed fully relativistic non-collinear LSDA+$U$+spin orbit (SO) calculations to check whether orbital angular momentum, $\vec{L}$, could have been unquenched by SO interaction in $t_{2g}$ orbitals, but we found no relevant changes in the magnetic moment ($\vec{L}\simeq 0$). 
 
\begin{figure}[htb]
\begin{center}
\includegraphics[clip=,height=7.7cm,width=0.5\textwidth]{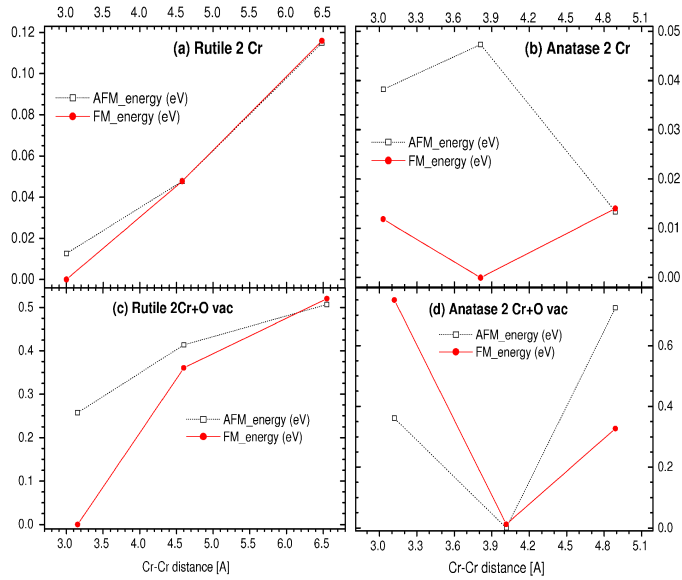}
\end{center}
\caption{(Color online) FM (filled circles) and AFM (open squares) LSDA+$U$ total energy vs interdistance of the two Cr impurities. The energy of the most stable configuration is set to zero. All values are in eV.}
\label{conf3}
\end{figure}

In Fig. \ref{conf3}, we plot the total LSDA+$U$ energy for the FM and the AFM configurations as a function of the interdistance of the two Cr impurities. All most stable configurations are FM, except for anatase with O vacancy, that is AFM (but nearly degenerate with the FM case: $\Delta E =  12$ meV).

{\it SE contributions to magnetic ground state. } 
We analyze here the stabilization of the FM ordering due to non-local spin-orbital correlations that are not included in the LSDA+$U$ starting point. It has to be noted that exchange interactions for Cr $t_{2g}$ electrons do not necessarily follow AFM Goodenough SE rules \cite{goodenough}, like in the case of Mn-$e_g$ electrons in perovskites. The sign of exchange interactions is affected by the $t_{2g}$ character of Cr electrons (contrary to $e_g$ electrons, they do not point towards ligand anions) and by the different connection of Cr ions in the two lattices, i.e., a $\sim 90\degree$ TM ion-anion-TM ion configuration in rutile, and a $\sim 180\degree$ TM ion-anion-TM ion configuration in anatase. In what follows, we provide the sign and we estimate the expected order of magnitude of the leading terms of SE interaction energies for the four configurations corresponding to Fig. \ref{conf3}. We find that only rutile with O vacancy has an AFM exchange energy, whose magnitude is however insufficient to overcome the FM LSDA+$U$ stabilization energy. All other three cases are instead characterized by FM corrections. In particular, the FM SE energy of anatase with O vacancy overcomes the 12 meV of the AFM LSDA+$U$ solution and leads to a FM ground state, as experimentally found.

Consider rutile at first: we refer to the model of \cite{prlmgtio,prbmgtio} for exchange of $t_{2g}$ electrons with a 90$\degree$ TM ion-anion-TM ion bond-angle. The exchange determined by the direct overlap of in-plane $t_{2g}$ orbitals (e.g., $d_{zx}$ orbitals in the $zx$ plane) is the leading term both with and without O vacancies.
In the former case, orbital degrees of freedom of Cr$^{3+}$ ion are frozen and the magnetic interaction is AFM, at odds with the case of 90$\degree$ TM ion-anion-TM ion of $e_g$ electrons, which is FM \cite{janisch}. The magnitude of this AFM exchange energy can be estimated as $E^{AFM}_r\simeq t^2/(U_2+J)\simeq 22$ meV, where $t\simeq 0.36$ eV is the hopping amplitude of the 2 Cr ions \cite{nota1} and $U_2\simeq U$ \cite{jphysc}. This AFM correction is however not sufficient to overcome the FM LSDA+$U$ contribution shown in Fig. 2(c). The case without O vacancies is different because Cr$^{4+}$ ions are characterized by one empty orbital and, because of the Coulomb repulsion, 2 Cr ions facing in the $xy$ plane have just one $d_{zx}$ orbital filled. This configuration favours FM-coupling \cite{prlmgtio}, with an energy gain of: $E^{FM}_r\simeq (J/U_2) t^2/(U_2-J)\simeq 5$ meV, slightly increasing the FM energy gap of Fig. 2(a). 

\begin{figure}[htb!]
\begin{center}
\includegraphics[clip=,height=5.2cm,width=0.48\textwidth]{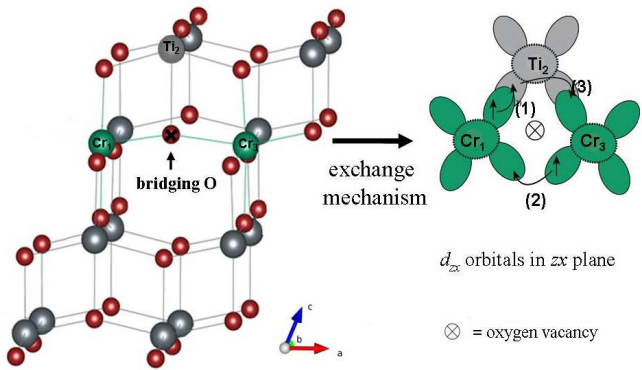}
\end{center}
\caption{(Color online) Schematic description of SE paths around the oxygen vacancy for charge-compensated anatase (see text). Ovals represent d$_{zx}$ orbitals.}
\label{conf4}
\end{figure}

For anatase, the change in topology ($\sim 180\degree$ TM ion-anion-TM ion configuration) also determines a change in the physical mechanism leading to exchange interactions.
The most interesting situation is that of anatase with O vacancies, as AFM and FM LSDA+$U$ solutions are nearly degenerate. Consider the hopping process around the O vacancy, through the $t_{2g}$-empty NN Ti-site. This is qualitatively described in Fig. \ref{conf4}: an electron from Cr$_1$ hops to Ti$_2$; then from Cr$_3$ to Cr$_1$; then from Ti$_2$ to Cr$_3$; the opposite path is also possible. This process is allowed only in case of FM coupling of Cr$^{3+}$ ions and is in competition with the two AFM processes where a) both electrons hop to Ti$_2$ with opposite spin; b) the two electrons directly hop from one Cr$^{3+}$ ion to the other, as for rutile, but with a much smaller hopping amplitude, $t_{13}$, as ions are much further. Notice that second-order hopping from Cr ions to NN Ti ion does not lead to effective exchange, as the energy gain does not depend on the magnetic coupling.
Quantitatively all these processes can be described through a Hamiltonian analogous to that of Ref. \cite{FeWO}, with intra-orbital Coulomb repulsion $U^{\rm Ti}_1$ at Ti-sites. 
We get a FM SE energy of: $E_{\rm FM}^{\rm SE}=-t^2t_{13}/\Delta_{\rm CT}^2\simeq -10$ meV. The AFM SE energies are: $E_{\rm AFM}^{\rm a}=-t^4/[\Delta_{\rm CT}^2(\Delta_{\rm CT}+U^{\rm Ti}_1)]\simeq -2$ meV and $E_{\rm AFM}^{\rm b}=-t_{13}^2/(U_2+6J)\simeq -0.6$ meV. Here $t_{13}\simeq 0.08$ eV, $U^{\rm Ti}_1\equiv U+2J\simeq 6.6$ eV and $\Delta_{\rm CT}\simeq 1$ eV is the charge-transfer energy to put an electron from a Cr site to a Ti site, as estimated from Fig. \ref{conf2}(f). Hopping parameters are scaled from Ref. \cite{prbmgtio} through the  law of Ref. \cite{harrison}, inversely proportional to the fifth-power interdistance.
There are two possible contributions for $E_{\rm FM}^{\rm SE}$ (the two opposite paths in the plane of Fig. \ref{conf4}), one for $E_{\rm AFM}^{\rm a}$ and 2 for $E_{\rm AFM}^{\rm b}$ and therefore the SE FM solution is stabilized by: $\Delta E_{\rm FM}^{\rm tot}\simeq 20-2-1 = 17$ meV, that overcomes the LSDA+$U$ AFM solution. We stress again the importance of the NNN condition in order to stabilize ferromagnetism: if the two Cr ions were NN, as in rutile, they would have been characterized by a strong AFM SE. This calculation shows that, although the F-center exchange is not active, Cr$^{3+}$ impurities contribute to magnetism via SE interactions, in keeping with experiments \cite{thin,prl95,prb73} and contrary to pure DFT results \cite{freeman}, or LSDA+U calculations (this work). 
  
It is important here to highlight that the FM SE stabilization energy strongly depends on $t$, $t_{13}$ and $\Delta_{\rm CT}$ parameters. In particular, increasing $t_{13}$ and decreasing $\Delta_{\rm CT}$ further stabilizes $E_{\rm FM}^{\rm SE}$. The opposite is true if $t_{13}$ decreases and $\Delta_{\rm CT}$ increase. All this suggests that, though they are not needed in our SE mechanism for FM, structural defects might change magnetic properties by changing the local values of $t_{13}$ and $\Delta_{\rm CT}$. In Ref. \cite{prl95,prb73} it is clearly stated that fast-grown samples, structurally defective, show a robust FM, contrary to non-defective samples. We suggest therefore to confirm or reject our proposed scenario by polarized EXAFS measurements with micro-beam at Cr K-edge, that can analyze the local environment of Cr atoms thereby showing whether $t_{13}$, the most critical of the three parameters, increases or decreases in fast-grown samples.


Finally, in the absence of O vacancies (Cr$^{4+}$), the previous mechanism is still active, but in competition with another SE interaction, mediated by the bridging O, where 2 electrons from the same $2p$-orbital at the bridging O-site move toward $e_g$ empty orbitals of Cr$^{4+}$ ions. This interaction is AFM because of the Hund's coupling with the underlying $t_{2g}$ orbitals, but its magnitude, for realistic values of the parameters, does not exceed 2 meV. Overall the LSDA+$U$ FM solution is again stabilized.

{\it $G_0W_0$@LSDA+$U$ results.} We double-check our zeroth-order LSDA+$U$ calculations through their GW corrections (at the $G_0 W_0$ level) by comparing the quasiparticle $G_0W_0$@LSDA+$U$ DOS with available photoemission experiments. As a general trend, $G_0W_0$ corrections strongly influence LSDA+$U$ DOS, with different contributions for spin-up and spin-down states. New empty impurity states appear, either isolated [Fig. 1(a), 1(b), 1(f)] or merged with the bottom of the conduction band (BCB) [Fig. 1(c), 1(d)].  

Our $G_0W_0$ results in samples with O vacancies (lowest LSDA+$U$ formation energy) provide filled t$_{2g}$ impurity states at 0.20 eV above the top of the valence band (TVB) for rutile and at 0.70 eV for anatase (FM solution, stabilized by SE). This allows us explaining with satisfying precision photoemission experiments \cite{thin} that found a state 0.3 eV (1.0 eV) from the TVB for doped rutile (anatase), thereby validating our approach. We remark that for anatase, had we evaluated the $G_0W_0$ corrections from the AFM LSDA+$U$ ground state (i.e., without SE corrections), we would have found two impurity states, at 1.30 and 2.00 eV, against experimental evidence \cite{thin}. 

A direct comparison of calculated band gaps with optical experiments is in principle invalidated by the neglect of excitonic effects in our approach. However, it is interesting to notice that calculated trends for $G_0W_0$ band gaps in doped and undoped rutile are in very good agreement with the experimental trend. In particular, we observe theoretically a big decrease in the band-gap of doped rutile (2.00-1.90 eV), compared to the undoped sample (3.70 eV), in keeping with the experimental trend of 1.98 eV for a doped sample \cite{biswas}, 3.21 eV for the undoped one \cite{rdholam}.

{\it Conclusions.} Our results show that the central role in both the induced magnetism and the position of impurity states is covered by the filling of the $t_{2g}$ levels of the dopant. In fact, $GW$@LSDA+$U$ DOS shows that the band gap is determined by $t_{2g}$ levels of the dopant and allows us to exclude the presence of impurity levels at the oxygen vacancy. The O vacancy is a F$^{++}$-center and the associated absence of electrons also implies that the F-center exchange, usually invoked to stabilize FM in these systems \cite{thin,coey}, can not be active in Cr-doped TiO$_2$. 
The role of $t_{2g}$-electrons for magnetism is highlighted by the new FM stabilization mechanism that we propose for anatase, based on the SE path of Fig. \ref{conf4}. Such a mechanism might be favored by oxygen vacancies, as the presence of a bridging oxygen opens a competing AFM SE channel, but it is also extremely sensitive to local structural distortions through hopping integrals. Actually, the relation of hopping and charge-transfer parameters with structural degrees of freedom (vacancies, structural defects, bond distance) becomes of crucial importance in such an approach and can indeed explain the enormous variety of results on magnetic ground states of a wide class of magnetically doped oxides. We suggested an experiment that can in principle confirm or reject our model. Methodologically, the present approach with LSDA+$U$ and SE corrections represents a general scheme that could be applied to other TMO in order to characterize magnetic ground states. In parallel, $G_0W_0$ calculations on top of LSDA+$U$(+SE) corrected results allow an independent check of the LSDA+$U$ ground state through a direct comparison of the quasiparticle spectrum with photoemission experiments.

\begin{acknowledgments}
F. Da Pieve and D. Lamoen gratefully acknowledge financial support from the University of Antwerp through the GOA project ``XANES meets ELNES''. The authors are grateful to Yann Pouillon for the technical
support with the build system of ABINIT. This work was supported by the Interuniversity Attraction Poles program (P6/42)-
Belgian State-Belgian Science Policy, the IWT-Vlaanderen ISIMADE project, the EU's 7th Framework
programme through the ETSF I3 e-Infrastructure project
(Grant Agreement No. 211956), the Communaut\'{e} francaise
de Belgique, through the Action de Recherche Concert\'{e}e
07/12-003 ``Nanosystemes hybrides metal-organiques'', 
the Walloon region Belgium (RW project N$^o$ 816849, WALL-ETSF) and
the FNRS through FRFC Project No. 2.4.589.09.F.
\end{acknowledgments}


\begin{thebibliography}{99}
\bibitem{science288} Y. Tokura and N. Nagaosa, Science {\bf 288}, 462 (2000)
\bibitem{secondo} P. Zhang, W. Luo, V. H. Crespi, M. L. Cohen and S. G. Louie, Phys. Rev. B {\bf 70}, 085108 (2004)
\bibitem{janesko} B. G. Janesko, T. M. Henderson, and G. E. Scuseria, Phys. Chem. Chem. Phys. {\bf 11}, 443 (2009)
\bibitem{droghetti} A. Droghetti, C.D. Pemmaraju, and S. Sanvito, Phys. Rev. B {\bf 78}, 140404(R) (2008)
\bibitem{method} H. Jiang, R.I. Gomez-Abal, P. Rinke and M. Scheffler, Phys. Rev. B {\bf 82}, 045108 (2010)
\bibitem{other1} H. Jiang, R.I. Gomez-Abal, P. Rinke and M. Scheffler, Phys. Rev. Lett. {\bf 102}, 126403 (2009)
\bibitem{other2} E. Kioupakis, P. Zhang, M.L. Cohen and S.G. Louie, Phys. Rev. B {\bf 77}, 155114 (2008)
\bibitem{science291} Y. Matsumoto, M. Murakami, T. Shono, T. Hasegawa, T. Fukumura, M. Kawasaki, P. Ahmet, T. Chikyow, S.-ya Koshihara and H. Koinuma, Science {\bf 291}, 854 (2001)
\bibitem{coey} J.M.D. Coey, M. Venkatesan and C. B. Fitzgerald, Nature Mater. {\bf 4}, 173 (2005)
\bibitem{freeman} Lin-Hui Ye and A. J. Freeman, Phys. Rev. B {\bf 73}, 081304(R) (2006)
\bibitem{jannotti} A. Janotti, J.B. Varley, P. Rinke, N. Umezawa, G. Kresse and C.G. van de Walle,  Phys. Rev. B {\bf 81}, 085212 (2010)
\bibitem{appl} Z. Wang, J. Tang, H. Zhang, V. Golub, L. Spinu, and L. D. Tung J. Appl. Phys. {\bf 95}, 7381 (2004)
\bibitem{appld}  X. Zhang, W. Wang, L. Li, Y. Cheng, X. Luo and H. Liu, J. Phys. D: Appl. Phys. {\bf 41}, 015005 (2008)
\bibitem{thin} J. Osterwalder, T. Droubay, T.C. Kaspar, J.R. Williams, C. Wang, S. Chambers, Thin Solid Films {\bf 484}, 289 (2005)
\bibitem{droub} T. Droubay, S. M. Heald, V. Shutthanandan, S. Thevuthasan, S. A. Chambers, and J. Osterwalder, J. Appl. Phys. {\bf 97}, 046103 (2005)
\bibitem{prl95}  T.C. Kaspar, S.M. Heald, C.M. Wang, J.D. Bryan, T. Droubay, V. Shutthanandan, S. Thevuthasan, D. E. McCready, A. J. Kellock, D. R. Gamelin, and S. A. Chambers, Phys. Rev. Lett. {\bf 95}, 217203 (2005) 
\bibitem{prb73} T.C. Kaspar {\it et al.}, Phys. Rev. B {\bf 73}, 155327 (2006) 
\bibitem{dival} C. Di Valentin, G. Pacchioni, H. Pnishi, A. Kudo, Chem. Phys. Lett. {\bf 469}, 166 (2009)
\bibitem{kyangmio} K.Yang, Y. Dai and B. Huang, Chem. Phys. Chem. {\bf 10}, 2329 (2009)
\bibitem{dmft} G. Kotliar, S. Y. Savrasov, K. Haule, V. S. Oudovenko, O. Parcollet, C. A. Marianetti, Rev. Mod. Phys. {\bf 78}, 865 (2006).
\bibitem{goodenough} J. Goodenough, Phys. Rev. {\bf 100}, 564 (1955)
\bibitem{abi} X. Gonze, B. Amadon, P.M. Anglade, J.M. Beuken {\it et al.}, Comp. Phys. Comm. {\bf 180}, 2582-2615 (2009)
\bibitem{godby} R. W. Godby and R.J. Needs, Phys. Rev. Lett. {\bf 62}, 1169 (1989)
\bibitem{bouquet} A. E. Bocquet, T. Mizokawa, K. Morikawa, A. Fujimori, S. R. Barman, K. Maiti, D. D. Sarma, Y. Tokura and M. Onoda, Phys. Rev. B {\bf 53}, 1161 (1996) 
\bibitem{serpone} N. Serpone, J. Phys. Chem. B {\bf 110}, 24287 (2006)
\bibitem{nhong} N.N. Hong, A. Ruyter, W. Prellier, and J. Sakai, Appl. Phys. Lett. {\bf 85} 6212 (2004)
\bibitem{prlmgtio} S. Di Matteo, G. Jackeli, C. Lacroix, N.B. Perkins, Phys. Rev. Lett. {\bf 93}, 077208 (2004)
\bibitem{prbmgtio} S. Di Matteo, G. Jackeli, N.B. Perkins, Phys. Rev. B {\bf 72}, 024431 (2005)
\bibitem{janisch} R. Janisch and N. A. Spaldin, Phys. Rev. B {\bf 73}, 035201 (2006) 
\bibitem{nota1} We consider hopping parameters of \cite{prbmgtio} and Refs. therein, with usual geometrical conditions of \cite{slater}.
\bibitem{slater} J.C. Slater and G.F. Koster, Phys. Rev. {\bf 94}, 1498 (1954)
\bibitem{jphysc} S. Di Matteo, N.B. Perkins and C.R. Natoli, J. Phys. Condens. Matter {\bf 14}, L37 (2002)
\bibitem{sangaletti} L.Sangaletti et al.; J.Phys.Condens. Matter 18, 7643-50 (2006)
\bibitem{FeWO} S. Di Matteo, G. Jackeli, N.B. Perkins, Phys. Rev. B {\bf 67}, 184427 (2003)
\bibitem{harrison} W. Harrison, {\it Electronic Structure and the Properties of Solids}, (Dover, New York, 1989)
\bibitem{biswas} S. Biswas, M. F. Hossain, T. Takahashi, Y. Kubota, A. Fujishima, Phys. Stat. Sol (a) {\bf 205}, 2023 (2008)
\bibitem{rdholam} R. Dholam,N. Patel,M. Adami,A. Miotello, Int. Journ. Hydr. En. {\bf 34}, 5337 (2009)

\end{thebibliography}

\end{document}